\documentclass[journal]{IEEEtran}

\ifCLASSINFOpdf
\else
\fi

\usepackage{amsmath}
\usepackage{cite}
\usepackage{amssymb}
\usepackage{graphicx}
\usepackage{enumerate}
\usepackage{comment,color}

\newtheorem{thm}{\bf Theorem}

\newtheorem{lem}{\bf Lemma}

\newtheorem{assum}{\bf Assumption}
\newtheorem{rem}{\bf Remark}

\newcommand*{\QEDB}{\hfill\ensuremath{\square}}

\newcommand{\R}{\mathbb{R}}

\newcommand{\NN}{\mathcal{N}}
\newcommand{\VV}{\mathcal{V}}

\newcommand{\WW}{\mathcal{W}}
\newcommand{\UU}{\mathcal{U}}

\newcommand{\col}{\mathrm{col}}
\newcommand{\avg}{\text{avg}}
\newcommand{\spn}{\mathrm{span}}
\newcommand{\fr}{\mathfrak{r}}

\newcommand{\bxa}{\bar x_{\text{avg}}}

\begin{document}

\title{Fully Distributed Resilient State Estimation\\ based on Distributed Median Solver}

\author{Jin~Gyu~Lee,~\IEEEmembership{Member,~IEEE,}
	Junsoo~Kim,~\IEEEmembership{Member,~IEEE,} \\
        and~Hyungbo~Shim,~\IEEEmembership{Senior Member,~IEEE,}%
        \thanks{This work was supported by Institute for Information \& communications Technology Promotion (IITP) grant funded by the Korea government (MSIT) (2014-0-00065, Resilient Cyber-Physical Systems Research), and by National Research Foundation of Korea (NRF) grant funded by the Korea government (Ministry of Science and ICT) (No. NRF-2017R1E1A1A03070342).
        This is a preprint of the following paper: Jin Gyu Lee, Junsoo Kim, and Hyungbo Shim, ``Fully distributed resilient state estimation based on distributed median solver,'' published in IEEE Transactions on Automatic Control, 2020, IEEE reproduced with permission of IEEE.
The final authenticated version is available online at: http://dx.doi.org/10.1109/TAC.2020.2989275}%
        \thanks{J. G. Lee is with the Control Group, Department of Engineering, University of Cambridge, United Kingdom. Email: {\tt jgl46@cam.ac.uk}. This author's work was done while he is with Seoul National University.}
        \thanks{J. Kim and H. Shim are with ASRI, Department of Electrical and Computer Engineering, Seoul National University, Korea. Email: {\tt kjs9044@cdsl.kr}, {\tt hshim@snu.ac.kr}}
        }
        
\maketitle

\begin{abstract}
In this paper, we present a scheme of fully distributed resilient state estimation for linear dynamical systems under sensor attacks.
The proposed state observer consists of a network of local observers, where each of them utilizes local measurements and information transmitted from the neighbors.
As a fully distributed scheme, it does not necessarily collect a majority of sensing data for the sake of attack identification, while the compromised sensors are eventually identified by the distributed network and excluded from the observers.
For this, the overall network (not the individual local observer) is assumed to have redundant sensors and assumed to be connected.
The proposed scheme is based on a novel design of a distributed median solver, which approximately recovers the median value of local estimates.
\end{abstract}

\begin{IEEEkeywords}
analytical redundancy, attack detection, attack resilience, cyber-physical systems, resilient state estimation, heterogeneous multi-agents, strong coupling, blended dynamics
\end{IEEEkeywords}

\section{Introduction}

As control systems are more connected and become vulnerable to cyber-sensor-attacks \cite{Teixeira15}, resilient state estimation problem has been posed.
Let the plant be given by
\begin{subequations}\label{eq:sys}
\begin{align}
\dot{x} &= Ax + B u,\label{eq:est_sys}\\
y &= Cx + a, \label{eq:output}
\end{align}
\end{subequations}
where $x\in\mathbb{R}^n$ is the state, $u\in\mathbb{R}^p$ is the input, $y\in\mathbb{R}^m$ is the output, and $a\in\mathbb{R}^m$ is the attack injected to the output sensor.
Throughout the paper, we suppose that the total of $m$ sensors is grouped into $N$ sensor banks, and the $i$-th sensor bank consists of $m_i$ sensors so that $\sum_{i=1}^N m_i = m$.
For convenience, each $i$-th block output $y_i \in \mathbb{R}^{m_i}$ of $y$ in \eqref{eq:output} is written as
$$y_i = C_i x + a_i\qquad i=1,2,\cdots,N,$$
where $C_i \in \mathbb{R}^{m_i \times n}$ is the $i$-th block rows of the output matrix $C$, and $a_i\in\mathbb{R}^{m_i}$ is the $i$-th block elements of the attack vector $a$.
Even though the attack signal $a$ can be arbitrarily designed by the adversary, there has been a common rationale that the number of corrupted sensors are limited due to the limited attack resources, e.g., \cite{Fawzi14}.
This rationale is formalized as the following assumption which implies up to $q$ sensor banks can be compromised out of $N$ sensor banks.

\begin{assum}\label{assum:q}
The attack signal $a_i(t)\in\R^{m_i}$ is identically zero for at least $N-q$ indices $i\in\NN := \{1,2,\cdots,N\}$; i.e.,
$\left| \{i \in \mathcal{N} : a_i(t) \equiv 0 \} \right| \ge N - q$
where $\left|\cdot\right|$ denotes the cardinality of a set. \QEDB
\end{assum}

In this paper, we consider the problem of {\it distributed resilient state estimation}.
The objective is to reconstruct the state $x$ in a distributed manner, in which the injection of sensor attack subject to Assumption~\ref{assum:q} is identified and excluded, so that it can not affect the state estimate.
More specifically, the problem is to construct state observers of the form
\begin{align}\label{eq:proposed}
\begin{split}
\dot{z}_i &= f_i^z(z_i, u, y_i), \\
\dot{\hat x}_i &= f^x_i(\hat x_i, z_i, u, \{\hat x_j\}_{j\in \mathcal{N}_i}),\quad i\in\mathcal{N}=\{1,2,\cdots,N\},
\end{split}
\end{align}
so that every estimate $\hat x_i$, $i\in\mathcal{N}$, recovers the true state $x$ even under attack.
The following descriptions specify the models of the observers and the adversary:
\begin{itemize}
\item The $i$-th local observer generates a partial estimate $z_i$ based on the information of the input $u$ and the $i$-th local output $y_i$, and then yields the estimate $\hat x_i$ for the state $x$ based on $z_i$, the input $u$, and the information of $\{\hat x_j\}_{j\in\mathcal{N}_i}$ that are transmitted from the neighbors, where $\mathcal{N}_i \subset \{1,2,\cdots,N\}$ is the set of nodes sending their information to the node $i$.\footnote{As the estimation is performed in a distributed way, a remark is made	that the plant is not assumed to be observable from a local output $y_i$, in general.}
\item The model of attack is subject to Assumption~\ref{assum:q}.
In particular, we consider an omniscient adversary\footnote{For example, the signal $a_i(t)$ for an attacked node can be generated, with the knowledge of all parameters and signals in \eqref{eq:sys} and \eqref{eq:proposed}.} and the non-zero $a_i(t)$ might have arbitrary values and is not assumed to be bounded.
\end{itemize}

In terms of the attack identification, it will be hopeless if {\em the majority} of the measured output data are compromised, since the compromised measurements may have arbitrary values.
As investigated in \cite{Pasqualetti13,Fawzi14}, every injection of $q$ attacks can be identified only when $q < N/2$, i.e., more than half of the sensor banks are guaranteed as un-compromised.
Especially, it is investigated in \cite{Fawzi14} that the resilient state estimation problem can be solved against every $q$ attacks only if the following condition of {\em $2q$-redundant observability} holds, which means \eqref{eq:est_sys} is observable from any $N-2q$ banks of measurements.

\begin{assum}\label{assum:red}
For any $\NN' \subset \mathcal{N} = \{1, \dots, N\}$ such that $|\NN'| = N-2q$, the matrix $C'$ which is a stacked matrix of $C_i$ for $i \in \NN'$ satisfies that the pair $(C', A)$ is observable. \QEDB
\end{assum}

Under $2q$-redundant observability, various solutions to the problem have been presented, most of which are essentially based on {\it majority voting}; each output measurement (or each estimate from the measurement) is to be compared with others and is identified as attack-free only when it coincides with the majority of measurements.
Since the problem is known to be NP-hard in general and combinatorial in nature \cite{Pasqualetti13}, reducing the computational complexity has been one of the most crucial issues for the existing results.
Many efforts have been made such as relaxation to convex optimizations \cite{Fawzi14,Shoukry16}, observer-based approaches \cite{Chong15,Chanhwa19,Junsoo19,Junsoo18}, and utilization of median functions \cite{jeon2016resilient,mitra2016secure}.
However, most of them are centralized schemes and they still require substantial computational efforts or resources as the number of sensors increases.

As distributed algorithms have been developed in various fields of study so as to divide a large computational problem into small-scale computations, distributed resilient state estimation (or distributed identification of sensor attack) has been tackled in recent years.
Related results can be found as in \cite{Pasqualetti15,mitra2016secure,Junsoo18,Chen18,mitra2019byzantine,an2019distributed}, but unfortunately, there is a gap between most of them and the true meaning of `fully distributed' solutions due to the lack of {\it distributed majority voting}.
Rather than presenting a majority voting algorithm in a distributed/cooperative manner, they assume that each local unit can collect a large number of measurements and then carry out the voting by itself.
As a result, in order to locally identify $q$ sensor attacks, they assumed that each local unit of sensors contains at least $2q+1$ sensors \cite{Pasqualetti15,Junsoo18}, or has at least $2q+1$ neighboring sensor nodes \cite{mitra2016secure,mitra2019byzantine,an2019distributed}.
On the other hand, a fully distributed attack identification scheme is made in \cite{Chen18}, but it is only for the case when the state has constant scalar value and it assumes only up to 30$\%$ of measurements can be compromised.

In this note, we present a scheme of fully distributed resilient state estimation for linear dynamical systems,
which also includes a way to identify sensor attacks in a fully distributed manner.
As a `fully distributed' solution, local observers in the proposed observer network, which takes the form of \eqref{eq:proposed}, do not necessarily collect a majority of sensor data for the sake of attack identification.
And thereby, there is no need for additional assumptions related to local identification.
In particular, compared with \cite{Pasqualetti15,Junsoo18}, there is no assumption that every local attack is identifiable from each local output $y_i$.
Especially, we do not assume that each $m_i$, the number of sensors in the $i$-th sensor bank, satisfies $m_i\ge 2q+1$.
With respect to the sensing redundancy against $q$ sensor attacks, we only assume $2q$-redundant observability of the overall system, which is also a necessary condition for resilient state estimation.
On the other hand, compared with \cite{mitra2016secure,mitra2019byzantine,an2019distributed}, each local observer does not need a majority of neighbors.
We only assume the network connectivity for the overall communication graph (see Assumption~\ref{assum:graph}), which is a necessary condition for distributed state estimation.

The algorithm that we use for a distributed majority voting is a distributed median solver developed in the next section.
The use of (centralized) median functions has been considered in \cite{jeon2016resilient} and \cite{mitra2016secure} to present a fast polynomial-time algorithm for resilient state estimation.
Taking advantage of the recent development of `blended dynamics' approach \cite{jglee}, the designed local observers approximately recover the median of the local state estimates in a distributed manner.
It will be seen that the median is a resilient estimate of the state even under up to $q$ corrupted sensor banks.
Thanks to one of the merits of blended dynamics approach \cite{jglee}, the proposed observer is different from \cite{Chen18} in that it allows the plug-and-play (initialization-free) operation; that is, the observer can perform resilient state estimation seamlessly even when some sensor banks or local observers intermittently join or leave the network as long as the proposed assumptions are maintained during the operation.

{\em Notation:}
Laplacian matrix $\mathcal{L} = [l_{ij}] \in \mathbb{R}^{N \times N}$ of a graph is defined as $\mathcal{L} := \mathcal{D} - \mathcal{A}$, where $\mathcal{A} = [\alpha_{ij}]$ is the adjacency matrix of the graph and $\mathcal{D}$ is the diagonal matrix whose diagonal entries is $\sum_{j = 1}^N \alpha_{ij}$.
By its construction, it contains at least one eigenvalue of zero, whose corresponding eigenvector is $1_N := [1,\dots,1]^T \in \mathbb{R}^N$, and all the other eigenvalues have nonnegative real parts.
For undirected graphs, the zero eigenvalue is simple if and only if the corresponding graph is connected.
For vectors or matrices $a$ and $b$, $\col(a,b) := [a^T,b^T]^T$.
For matrices $A_1, \dots, A_k$, we denote by $\text{diag}(A_1, \dots, A_k)$ the block diagonal matrix.
The operation defined by the symbol $\otimes$ is the Kronecker product.
The maximum norm of a vector $x$ is defined by $\|x\|_\infty := \max_i |x_i|$, and the Euclidean norm is denoted by $\|x\| := \sqrt{x^Tx}$.
The induced Euclidean norm of a matrix $A$ is written by $\|A\|$.
For a set $\Xi$, $\| x \|_{\Xi}$ denotes the distance between the vector $x$ and $\Xi$, i.e., $\| x \|_{\Xi} := \inf_{y \in \Xi} \| x - y\|$.
An interval $[a,b]$ of real numbers $a$ and $b$ implies $\{ x : a \le x \le b \}$.
For a set ${\mathcal Z}$, its cardinality is denoted by $|{\mathcal Z}|$.
The function $\text{sgn}: \R \to \R$ denotes the signum function defined as $\text{sgn}(s) = s/|s|$ for non-zero $s$, and $\text{sgn}(s) = 0$ for $s = 0$.
For a real number $x$, rounding up and down of $x$ is denoted by $\lceil x \rceil$ and $\lfloor x \rfloor$, respectively.
In this paper, any solution of a discontinuous dynamical system is considered as a Filippov solution, any adjacency element $\alpha_{ij}$ is $0$ or $1$, and all positive (semi)definite matrices are symmetric.

\section{Distributed Median Solver}\label{sec:pre}

Despite the general problem we are dealing with in this paper, let us consider for a moment a simplified problem of estimating a constant scalar value given the corrupted set of constant scalar outputs where the attack is assumed to be also a constant, i.e., $x \in \mathbb{R}$, $A = 0$, $B = 0$, and $y_i \in \mathbb{R}$ for all $i \in \mathcal{N}$ in \eqref{eq:sys}.
Then, Assumption~\ref{assum:red} implies that there is more than $2q+1$ indices $i$ such that $C_i$ is non-zero.
Let $s_i$ be $1$ when $C_i$ is non-zero and $0$ otherwise.
Since by Assumption~\ref{assum:q}, there is at most $q$ indices $i$ such that $a_i$ is non-zero, a simple solution to obtain a resilient estimate of a constant scalar value $x$ is to take a majority vote among the local estimates $z_i := y_i/C_i$ for all the indices $i$ such that $s_i =1$.
One particular tool which makes this happen is a median operation, and by constructing a distributed median solver, we can thus solve the distributed resilient state estimation problem for this special case.
Thus, in this section, we propose a distributed median solver, and in the later section, we will see how this can be extended to cover the general problem illustrated in the Introduction.

In this section, for a collection $\mathcal{Z}$ of real numbers $z_i$, $i=1, 2, \cdots, N$ associated with a collection $\mathcal{S}$ of indicators $s_i \in \{ 0, 1\}$, $i = 1, 2, \cdots, N$ (with at least one index $i$ such that $s_i = 1$), a median among the indicated values ($z_i$ for $i$ such that $s_i = 1$) is defined as a real number that belongs to the set
$$\mathcal{M}_{\mathcal{Z}}^{\mathcal{S}} = \begin{cases} \{z_{(S+1)/2}^s\}, &\mbox{ if $S$ is odd}, \\
[z_{S/2}^s, z_{S/2 + 1}^s], &\mbox{ if $S$ is even}, \end{cases}$$
where $S$ is the number of indicated values, i.e., $S := | \{i  \in \mathcal{N} : s_i = 1\}|$, and $z_i^s$'s are the elements of the set of indicated values $\{ z_i : s_i = 1 \}$ with its index being rearranged (sorted) such that
$$z_1^s \le z_2^s \le \cdots \le z_S^s.$$
With the help of this relaxed definition of the median, finding a median $x$ of ${\mathcal Z}$ associated with $\mathcal{S}$ becomes solving a simple optimization problem
$$\text{minimize}_x \,\, \begin{matrix} \sum_{i=1}^N\end{matrix}s_i|z_i - x|.$$
Then, the gradient descent algorithm given by
\begin{align}\label{eq:da_dms}
\dot{\hat{x}} = \begin{matrix}\sum_{i=1}^N \end{matrix} s_i\text{sgn}(z_i - \hat{x})
\end{align}
will solve this minimization problem.
In particular, the solution $\hat{x}$ satisfies
$$\lim_{t \to \infty} \|\hat{x}(t)\|_{\mathcal{M}_{\mathcal{Z}}^{\mathcal{S}}} = 0.$$

Motivated by this, we propose a distributed median solver, whose individual dynamics of the agent $i$ uses the information of $z_i$ and $s_i$ only: 
\begin{align}\label{eq:dms}
\dot{x}_i =  \, s_i\text{sgn}(z_i - x_i) + \gamma \begin{matrix}\sum_{j \in \mathcal{N}_i}\end{matrix} (x_j - x_i), \quad i \in \mathcal{N},
\end{align}
where $\gamma > 0$ is a design parameter.

Now, under a mild assumption on the graph, the algorithm \eqref{eq:dms} finds a median approximately by exchanging their states $x_i$ only (not $z_i$ nor $s_i$).

\begin{assum}\label{assum:graph}
The graph is undirected and connected. \QEDB
\end{assum}

\begin{thm}\label{thm:sup_med}
Let Assumption \ref{assum:graph} hold.
Then, for each $\gamma > 0$, the solution to \eqref{eq:dms} from any initial condition $x_i(0) \in \mathbb{R}$, $i \in \NN$, exists for all $t \ge 0$ and satisfies
$$\limsup_{t \to \infty} \|x_i(t)\|_{\mathcal{M}_{\mathcal{Z}}^{\mathcal{S}}} \le \frac{2\sqrt{N}}{\gamma\lambda_2(\mathcal{L})}, \qquad \forall i \in \mathcal{N},$$
where $\mathcal{Z}$ and $\mathcal{S}$ are a given set of numbers $\{z_i\}_{i \in \mathcal{N}}$ and a given set of indicators $\{s_i\}_{i \in \mathcal{N}}$ respectively (with at least one index~$i$ such that $s_i = 1$), and $\lambda_2(\mathcal{L})$ is the algebraic connectivity of the graph (i.e., the second smallest eigenvalue of the Laplacian matrix $\mathcal{L}$ that represents the graph). \QEDB
\end{thm}

\begin{rem}
The insight behind the proposed distributed median solver \eqref{eq:dms} comes from the so-called `blended dynamics' approach \cite{jglee}.
In this approach, the behavior of heterogeneous multi-agent systems
$$\dot{x}_i = f_i(t,x_i) + \gamma \begin{matrix}\sum_{j \in \mathcal{N}_i} \end{matrix}(x_j - x_i), \quad i \in \mathcal{N},$$
with large coupling gain $\gamma$ is approximately estimated by the behavior of the blended dynamics defined by
$$\dot{\hat{x}} = (1/N)\begin{matrix}\sum_{i=1}^N\end{matrix} f_i(t,\hat{x}).$$
In our case, the blended dynamics of \eqref{eq:dms} is obtained as the gradient descent algorithm \eqref{eq:da_dms} with $1/N$ scaling in time. \QEDB
\end{rem}

\begin{IEEEproof}
The proof is provided in Appendix~\ref{app:proof_dms}.
\end{IEEEproof}

\begin{rem}
The algebraic connectivity $\lambda_2(\mathcal{L})$ depends on both the topology of the graph and the number $N$ of the nodes.
For example, $\lambda_2(\mathcal{L}) = 2(1 - \cos(2\pi/N))$ for the ring network, so that it decreases as $N$ increases.
On the other hand, for the all-to-all networks, $\lambda_2(\mathcal{L})$ is the same as the number $N$ \cite{fiedler1973algebraic}.
Therefore, if the network graph is all-to-all, the increase of the node number $N$ actually improves the steady-state error $2\sqrt{N}/(\gamma \lambda_2(\mathcal{L})) = 2/(\gamma \sqrt{N})$. \QEDB
\end{rem}

\section{Distributed Resilient State Estimation}\label{sec:drse}

\subsection{Applicable Class of Systems}\label{sec:ACS}
 
In this section, we return to our original problem illustrated in the Introduction, which considers distributed resilient estimation of a high-dimensional time-varying vector $x(t) \in \mathbb{R}^n$.
The basic idea in this section is to find a unified coordinate transformation so that we can perform element-wise median operation among the partial estimates, generated based on each output, to obtain a resilient estimate of the whole vector $x(t)$.

For this, with $\UU_i$ being the unobservable subspace of the pair $(C_i,A)$ for each $i \in \NN$, we require the following technical assumption.

\begin{assum}\label{assum:class}
There exists a basis $\{ v_1, \dots, v_n \}$ of $\R^n$ such that every $\UU_i$, $i \in \NN$, is a span of a subset of the basis.

\QEDB
\end{assum}

The key to the assumption is that the {\em same} basis is used to express all different subspaces $\UU_i$.
To see how strong/weak this assumption is, we refer to Appendix \ref{app:ASC} where we list a few sufficient conditions for Assumption \ref{assum:class}.
We emphasize, from Appendix \ref{app:ASC}, that Assumption \ref{assum:class} holds if the characteristic polynomial and the minimal polynomial of~$A$ are the same.

Now let us define an indicator $s_i^{l}$ such that
$$s_i^{l} = \begin{cases} 1, &\text{if } v_l \notin \mathcal{U}_i, \\ 0, &\text{if } v_l \in \mathcal{U}_i. \end{cases}$$
Equivalently, we have $s_i^l = 1$ if the value $w_l^Tx$ is observable from the sensor bank $y_i = C_ix$, where $w_l \in \mathbb{R}^n$ is such that 
$$\begin{bmatrix} w_1 & \cdots & w_n \end{bmatrix}^T \begin{bmatrix} v_1 & \cdots & v_n \end{bmatrix} = I_n.$$
Then, as seen in the following lemma, the $2q$-redundant observability (Assumption~\ref{assum:red}) implies that each $w_l^Tx$, $l=1,\cdots,n$, is observable from at least $2q+1$ sensor banks.

\begin{lem}\label{lem:s}
Let Assumptions \ref{assum:red} and \ref{assum:class} hold.
Then, for each $l = 1, \dots, n$, it holds that $|\{ i \in \NN : s_i^l = 1 \}| \ge 2q+1$. \QEDB
\end{lem}

\begin{IEEEproof}
Assume that this is not true.
Then, there exist $l \in \{1, \dots, n\}$ and $\mathcal{N}' \subset \mathcal{N}$ such that $|\mathcal{N}'| = N - 2q$ and $v_l \in \mathcal{U}_i$ for all $i \in \mathcal{N}'$.
Now, this contradicts Assumption \ref{assum:red}.
\end{IEEEproof}

It is then seen that, under Assumptions~\ref{assum:red} and \ref{assum:class}, resilient state estimation against up to $q$ sensor attacks is obtained by element-wise majority voting.
In particular, if $s_i^{l} = 1$, then one should be able to design an estimator that yields $z_i^l$, which is the estimate of $w_l^Tx$, from $y_i = C_i x$.
If we collect all $z_i^{l}$, there are more than or equal to $2q+1$ estimates due to Lemma \ref{lem:s}.
This means that, even if up to $q$ sensor banks are corrupted by adversaries so that up to $q$ estimates become untrustful, there are still at least $q+1$ trustful estimates.
Therefore, a majority vote from all estimates yields a trustful estimate, and a simple way to do the majority vote is to take the median of all estimate candidates.
Then, by collecting these resilient estimates for all $l = 1, 2, \cdots, n$, each agent achieves resilient estimation of the whole state $x(t)$.
Note that the state observer that uses $y_i$ can be installed at the $i$-th sensor bank, and therefore, the estimate $z_i^{l}$ can be obtained locally.
The forthcoming subsections show how this can be done precisely.

\begin{rem}
For interested readers, we present another explanation for the role of Assumption~\ref{assum:class}.
As noted in the Introduction, $2q$-redundant observability is a necessary condition for resilient state estimation.
On the other hand, a well-used sufficient condition is the null-space property \cite[Proposition 6]{Fawzi14}.
It is well-known that under the null-space property, it is possible to avoid solving NP-hard problems by equivalently converting the $l_0$ minimization problem into an $l_1$ minimization problem.
By recalling that the median operation illustrated in Section~\ref{sec:pre} was considered as an $l_1$ minimization problem, we might think that Assumption~\ref{assum:class} provides a connection between the necessary and the sufficient condition.
In fact, unlike the terminology, the null-space property is not the condition about the null-space (or in the context of this paper, about the unobservable subspace $\UU_i$) and it depends on the specific choices of matrices even if those matrices have the same null-space.
Now, the role of Assumption~\ref{assum:class} is to guarantee under the necessary condition ($2q$-redundant observability) that we can always find a set of matrices that satisfies the null-space property.
Therefore, under Assumptions~\ref{assum:red} and \ref{assum:class}, we can solve an equivalent $l_1$ minimization problem (which is to take the median) to obtain a resilient estimate.
A typical choice of matrices are given in the next subsection, and the null-space property directly follows from Lemma~\ref{lem:s}.
In this regard, it seems further studies are required for the gap between the necessary and the sufficient condition. \QEDB
\end{rem}

\subsection{Proposed Distributed Resilient State Observer}

Putting all the discussions so far together, our design of the distributed state observer \eqref{eq:proposed} is proposed.
With the unobservable subspace $\UU_i$ of $(C_i,A)$ and the basis $\{v_1,\dots,v_n\}$ of Assumption \ref{assum:class}, let $\VV_i$ be an $n \times o_i$ matrix, where $o_i = n - \dim(\UU_i)$, whose columns are $v_l$ with $v_l \notin \mathcal{U}_i$, so that the columns of $\VV_i$ are a basis of the observable (quotient) subspace of $(C_i, A)$.
Moreover, let $\WW_i$ be an $o_i \times n$ matrix, whose rows are $w_l^T$ with $s_i^l = 1$.
Then, $\WW_i\VV_i = I_{o_i}$ and we obtain a Kalman observability decomposition as
$$\begin{bmatrix} \WW_i \\ \tilde{\WW}_i\end{bmatrix} A \begin{bmatrix} \VV_i & \tilde{\VV}_i\end{bmatrix} = \begin{bmatrix} * & 0 \\ * & * \end{bmatrix}, \quad C_i\begin{bmatrix} \VV_i & \tilde{\VV}_i \end{bmatrix} = \begin{bmatrix} * & 0 \end{bmatrix},$$
where $\tilde{\WW}_i \in \mathbb{R}^{(n-o_i) \times n}$ and $\tilde{\VV}_i \in \mathbb{R}^{n \times (n-o_i)}$ consist of $w_l^T$ and $v_l$ with $s_i^l = 0$, respectively such that $\tilde{\WW}_i\tilde{\VV}_i = I_{n-o_i}$.
Design an observer gain matrix $L_i \in \R^{o_i \times m_i}$ such that $\WW_i A \VV_i - L_i C_i \VV_i$ is Hurwitz, which is possible since $(C_i\VV_i, \WW_iA\VV_i)$ is observable.
Hence, a partial observer for each $i$-th sensor bank becomes
\begin{align}\label{eq:z_i}
\dot z_i &= \WW_i (A \VV_i z_i + B u) + L_i ( y_i - C_i \VV_i z_i ) \in \mathbb{R}^{o_i}.
\end{align}
This observer estimates the state as much as possible from the available information $y_i$.
In particular, when the $i$-th sensor bank is attack free, i.e., $a_i \equiv 0$, we can estimate $w_l^Tx$, as a component of $z_i$ in \eqref{eq:z_i}, for all $l$ such that $s_i^l = 1$.
However, the estimate of $w_l^Tx$ from the $i$-th sensor bank may be corrupted if $y_i$ is corrupted by the attack signal $a_i$.
Nevertheless, by recalling from Lemma~\ref{lem:s} that, for each $l = 1, 2, \cdots, n$, there are at least $2q+1$ sensor banks from each of which we can estimate $w_l^Tx$ through the partial observer \eqref{eq:z_i}, and that there are up to $q$ sensor attack, it is left to take the median out of \emph{all} the estimates for each $w_l^Tx$, to neglect the corrupted estimates.

Here we note that $w_l^Tx(t)$ to be estimated is time-varying, and so, the distributed median solver of the form \eqref{eq:dms} may not efficiently track $w_l^Tx(t)$.
Therefore, inspired by the internal model principle, our idea is to embed the model for the state $x(t)$ into the distributed median solver as
\begin{align} \label{eq:x_i}
\dot {\hat x}_i &= A \hat x_i + B u + \kappa \sum_{l = 1}^n s_i^l \text{sgn}(w_l^T\VV_iz_i - w_l^T\hat{x}_i) v_l \nonumber \\
&\quad\quad\quad\quad\quad\quad\quad\quad + \kappa\gamma\begin{matrix}\sum_{j \in \mathcal{N}_i}\end{matrix} (\hat{x}_j - \hat{x}_i) \,\,\in \mathbb{R}^n
\end{align}
where positive gains $\kappa$ and $\gamma$ are design parameters.
It is noted that, in \eqref{eq:x_i}, the correction of $\hat x_i(t)$ is performed by the last two summation terms.
In the first summation, the value $w_l^T\VV_iz_i$ is identically zero for some $l \in \{1,\dots,n\}$, which means that we cannot estimate $w_l^Tx$ from the output $y_i$.
In order not to perturb the $\hat x_i$-dynamics, the indicator $s_i^l$ becomes zero as well in this case.
Then, the correction, in this case, is actually performed by the second summation; that is, the unobservable components of $\hat x_i$ is compensated by the estimates of the neighboring agents.

The next theorem shows that, for sufficiently large $\kappa$ and $\gamma$, the distributed algorithm \eqref{eq:z_i} and \eqref{eq:x_i} achieves a resilient estimation of $x$ with arbitrary precision.

\begin{thm}\label{thm:suff}
Let Assumptions \ref{assum:q}, \ref{assum:red}, \ref{assum:graph}, and \ref{assum:class} hold.
Then, for each compact set $K \subset \mathbb{R}^{n + \sum_{i=1}^N (o_i + n)}$ and $\eta > 0$, there exist $\kappa^*$ and $\gamma^*$ such that, for each $\kappa > \kappa^*$, $\gamma > \gamma^*$, and ${\rm col}(x(0), z_1(0), \hat{x}_1(0), \dots, z_N(0), \hat{x}_N(0)) \in K$, the solution to \eqref{eq:sys}, \eqref{eq:z_i}, and \eqref{eq:x_i} exists for all $t \ge 0$, and satisfies
\begin{align*}
\limsup_{t \to \infty} \|\hat{x}_i(t) - x(t)\|_\infty &\le  \eta, 
\end{align*}
for all $i \in \mathcal{N}$. \QEDB
\end{thm}

\begin{IEEEproof}
Define the error variables as $\bar{x}_i := \mathcal{W}(\hat{x}_i - x) \in \mathbb{R}^n$, where $\mathcal{W} = {\rm col}(w_1^T, \dots, w_n^T)$.
Then, the error dynamics is given for each $i \in \mathcal{N}$ as ($\bar{x}_i = {\rm col}(\bar{x}_i^1, \dots, \bar{x}_i^n)$)
\begin{equation}\label{eq:key}
\dot{\bar{x}}_i = \bar{A}\bar{x}_i + \kappa\begin{bmatrix} s_i^1 \text{sgn}(p_i^1 - \bar{x}_i^1) \\ \vdots \\ s_i^n \text{sgn}(p_i^n - \bar{x}_i^n) \end{bmatrix} + \kappa\gamma\sum_{j \in\mathcal{N}_i} (\bar{x}_j - \bar{x}_i)
\end{equation}
where $p_i = {\rm col}(p_i^1, \dots, p_i^n) := \mathcal{W}\VV_iz_i - \mathcal{W}x \in \mathbb{R}^n$, $\bar{A} = \mathcal{W}A\mathcal{V}$, and $\mathcal{V} = [v_1, \dots, v_n]$.
Let a matrix $R \in \mathbb{R}^{N \times (N-1)}$ be a matrix whose columns are orthogonal unit vectors such that each column is perpendicular to $1_N$.
Then, by defining
$\bar{x}_{\text{avg}} = {\rm col}(\bar{x}_{\text{avg}}^1, \dots, \bar{x}_{\text{avg}}^n) := (1/N) \sum_{i=1}^N \bar{x}_i$ and $\tilde{x} := (R^T \otimes I_n)\,\,{\rm col}(\bar{x}_1, \dots, \bar{x}_N)$, we have $\bar{x}_i = \bar{x}_{\text{avg}} + (\fr_i^T \otimes I_n)\tilde{x}$ where $\fr_i^T$ is the $i$-th row of $R$.
Therefore, it follows that
\begin{align*}
\dot{\bar{x}}_{\text{avg}} &= \bar{A}\bar{x}_{\text{avg}} + \frac{\kappa}{N} \sum_{i=1}^N \begin{bmatrix} s_i^1 \text{sgn}(p_i^1(t) - \bar{x}_{\text{avg}}^1 - (\fr_i^T \otimes e_1^T)\tilde{x}) \\ \vdots \\ s_i^n \text{sgn}(p_i^n(t) - \bar{x}_{\text{avg}}^n  - (\fr_i^T \otimes e_n^T)\tilde{x}) \end{bmatrix}\!\!, 
\end{align*}
\begin{align*}
\dot{\tilde{x}} &= (I_{N-1} \otimes \bar{A})\tilde{x} - \kappa\gamma (Q \otimes I_n )\tilde{x} \\
&\,\,\,\,\,+ \kappa (R^T \otimes I_n)\begin{bmatrix} s_1^1\text{sgn}(p_1^1(t) - \bar{x}_{\text{avg}}^1 - (\fr_1^T \otimes e_1^T)\tilde{x}) \\ \vdots \\ s_1^n \text{sgn}(p_1^n(t) - \bar{x}_{\text{avg}}^n - (\fr_1^T \otimes e_n^T)\tilde{x}) \\ \vdots \\ s_N^n \text{sgn}(p_N^n(t) - \bar{x}_{\text{avg}}^n - (\fr_N^T \otimes e_n^T) \tilde{x}) \end{bmatrix}\!\!,
\end{align*}
where the matrix $Q := R^T\mathcal{L}R$ is positive definite and $e_l$ is the elementary vector, i.e., the $l$-th element of $e_l$ is one and all other elements of $e_l$ are zero.

Now, with $W(t) := \|\tilde x(t)\|$, it is seen that, when $W > 0$, 
\begin{align}\label{eq:dyn_W}
&\dot{W} = \frac{\tilde{x}^T\dot{\tilde{x}} + \dot{\tilde{x}}^T\tilde{x}} {2\sqrt{\tilde{x}^T\tilde{x}}} \nonumber \\
&\le \frac{1}{2W}\tilde{x}^T(I_{N-1} \otimes (\bar{A} + \bar{A}^T))\tilde{x} - \kappa\gamma\lambda_{\text{min}}(Q)W +\kappa\sqrt{Nn} \nonumber \\
&\le \kappa \sqrt{Nn} - (\kappa\gamma\lambda_2(\mathcal{L}) - \|\bar{A}\|)W.
\end{align}
Then, with $\delta_\eta := \eta/(3\|\mathcal{V}\|_\infty)$, it can be shown that, if
$$\gamma > \frac{2 \kappa N\sqrt{N}n^2 + \delta_\eta \|\bar A\|}{\kappa \delta_\eta \lambda_2(\mathcal{L})} =: \bar \gamma(\kappa),$$
then we have
$$\dot W \le - \frac{\kappa N\sqrt{N}n^2}{\delta_\eta} W < 0 \quad \text{when} \quad W \ge \frac{\delta_\eta}{Nn\sqrt{n}} =: \delta_\eta'.$$
Hence, since all the initial conditions belong to the compact set $K$, there are a bound $B_w$ and a time $T_w$ such that $W(t) = \|\tilde x(t)\| \le B_w$ for all $t \ge 0$, and $W(t) = \|\tilde x(t)\| \le \delta_\eta'$ for all $t \ge T_w$.

On the other hand, if there is no attack, then it is clear that $\lim_{t \to \infty}p_i^l(t)=0$ for any pair $(i,l)$ such that $s_i^l=1$.
Since the initial conditions belong to the compact set $K$, there are a bound $B_p$ and a time $T_p$ such that $|p_i^l(t)| \le B_p$ for all $t \ge 0$ and $|p_i^l(t)| \le \delta_\eta'$ for all $t \ge T_p$.

Now, define a set
$$\theta(\bar x_{\text{avg}}) := \{ l \in \{1,\cdots , n\} : |\bar x_{\text{avg}}^l| \ge \|\bar x_{\text{avg}}\|_\infty/(Nn) \}$$
which is non-empty because there is $l^*$ such that $\|\bxa\|_\infty = |\bxa^{l^*}|$ by definition of the infinity norm.
Suppose that $\bxa$ is given.
For each $l \in \theta(\bxa)$, there are at least $q+1$ indices of $i$ such that $s_i^l=1$ and $a_i(t) \equiv 0$ by Lemma~\ref{lem:s} and Assumption~\ref{assum:q}.

\smallskip

{\em Claim:} For such pairs $(i,l)$ (i.e., $l \in \theta(\bxa)$, $s_i^l = 1$, and $a_i(t) \equiv 0$), if
\begin{equation}\label{eq:claim}
|p_i^l - (\fr_i^T \otimes e_l^T) \tilde x| < \frac{\|\bar x_{\text{avg}}\|_\infty}{Nn},
\end{equation}
then, with $V(t) = \|\bxa(t)\|$,
$$\dot V \le \|\bar A\| V - \frac{\kappa}{Nn\sqrt{n}}.$$
{\em Proof of Claim:}
Since $l \in \theta(\bar{x}_{\text{avg}})$, we have that $|p_i^l - (\fr_i^T \otimes e_l^T) \tilde x| < |\bxa^l|$, and thus,
\begin{align}\label{eq:maj_vote}
\sum_{i=1}^N s_i^l \bar{x}_{\text{avg}}^l\text{sgn}(p_i^l - \bar{x}_{\text{avg}}^l - (\fr_i^T \otimes e_l^T)\tilde{x}) \le -|\bar{x}_{\text{avg}}^l|
\end{align}
because the number of $i$'s such that $s_i^l=1$ and $a_i \equiv 0$ is at least one more than the number of $i$'s such that $s_i^l=1$ and $a_i \not \equiv 0$.
It follows from \eqref{eq:maj_vote} that
\begin{align*}
\dot{V} &\le \frac{1}{2V}\bar{x}_{\text{avg}}^T(\bar{A} + \bar{A}^T)\bar{x}_{\text{avg}} \\
&\quad  + \frac{1}{V}\frac{\kappa}{N}\sum_{l=1}^n\sum_{i=1}^N s_i^l \bar{x}_{\text{avg}}^l \text{sgn}(p_i^l - \bar{x}_{\text{avg}}^l - (\fr_i^T\otimes e_l^T)\tilde{x}) \\
&\le \|\bar{A}\|V + \frac{1}{V} \frac{\kappa}{N} \sum_{l \not\in \theta} \sum_{i=1}^N \frac{1}{Nn} \|\bar{x}_{\text{avg}}\|_\infty \\ 
& + \frac{1}{V}\frac{\kappa}{N} \sum_{l \in \theta} \sum_{i=1}^N s_i^l\bar{x}_{\text{avg}}^l \text{sgn}(p_i^l - \bar{x}_{\text{avg}}^l - (\fr_i^T\otimes e_l^T)\tilde{x})  \\
&\le \|\bar{A}\| V + \frac{1}{V} \frac{\kappa}{N} \frac{n-|\theta|}{n} \|\bar{x}_{\text{avg}}\|_\infty - \frac{1}{V}\frac{\kappa}{N}\sum_{l \in \theta} |\bar{x}_{\text{avg}}^l| \\
&\le \|\bar{A}\| V + \frac{1}{V} \frac{\kappa}{N} \frac{n-1}{n} \|\bar{x}_{\text{avg}}\|_\infty - \frac{1}{V}\frac{\kappa}{N}|\bar{x}_{\text{avg}}^{l^*}| \\
&\le \|\bar{A}\|V - \frac{1}{V}\frac{\kappa}{Nn}\|\bar{x}_{\text{avg}}\|_\infty
\le \|\bar{A}\|V - \frac{\kappa}{Nn\sqrt{n}}
\end{align*}
where $l^*$ is such that $|\bxa^{l^*}| = \|\bxa\|_\infty$ (both $l^*$ and $\theta$ depend on $\bxa$), and the last inequality follows from that $\|\bxa\| \le \sqrt{n} \|\bxa\|_\infty$.
This completes the proof of the Claim.

\smallskip

Let $\bar M$ be a constant such that $V(0) \le \bar M$ and $B_p + B_w \le \bar M/(Nn\sqrt{n})$, which exists since all the initial conditions belong to the compact set $K$.
Then, whenever $V > \bar M$, we get for each pair $(i, l)$ which satisfies $s_i^l = 1$ and $a_i \equiv 0$,
$$|p_i^l(t) - (\fr_i^T \otimes e_l^T)\tilde{x}| \le \frac{V}{Nn\sqrt{n}} = \frac{\|\bar{x}_{\text{avg}}\|}{Nn\sqrt{n}} \le \frac{\|\bar{x}_{\text{avg}}\|_\infty}{Nn}.$$
Then, by the Claim, it is seen that $\dot V \le \|\bar A\|V$, and thus, $V(t) \le \bar M \exp(\|\bar A\|t)$ for all $t \ge 0$.

Let $\kappa^* := Nn\sqrt{n}\|\bar{A}\|\bar{M}\text{exp}(\|\bar{A}\|T)$ where $T := \max\{T_w,T_p\}$.
Then, for any $\kappa > \kappa^*$ and $\gamma > \bar \gamma(\kappa^*) =: \gamma^*$ (where $\bar \gamma(\cdot)$ is a decreasing function so that $\gamma > \bar \gamma(\kappa)$), we have that $|p_i^l(t) - (\fr_i^T \otimes e_l^T) \tilde x(t)| \le |p_i^l(t)| + \|\tilde x(t)\| \le 2 \delta_\eta' = 2\delta_\eta/(Nn\sqrt{n})$ after the time $T$ for such pairs $(i,l)$ that $s_i^l=1$ and $a_i \equiv 0$.
Moreover, it follows from the Claim that
\begin{align*}
\dot V \le \|\bar A\| \bar M \exp(\|\bar A\|T) - \frac{\kappa}{Nn\sqrt{n}} < 0 \quad \text{if} \quad \|\bar x_{\text{avg}}\|_\infty > \frac{2\delta_\eta}{\sqrt{n}}.
\end{align*}
This implies that
\begin{align*}
\limsup_{t \to \infty} \|\bar{x}_{\text{avg}}(t)\|_\infty &
\le \limsup_{t \to \infty} V(t) \le 2\delta_\eta,
\end{align*}
and finally implies that, for all $i \in \mathcal{N}$,
\begin{align*}
&\limsup_{t \to \infty} \|\hat{x}_i(t) - x(t)\|_\infty = \limsup_{t \to \infty} \|\mathcal{V}\bar{x}_i(t)\|_\infty  \\
&\le \|\VV\|_\infty \limsup_{t \to \infty} \| \bar{x}_{\text{avg}}(t) + (\fr_i^T \otimes I_n)\tilde{x} \|_\infty \le 3\|\mathcal{V}\|_\infty\delta_\eta = \eta.
\end{align*}
This concludes the proof.
\end{IEEEproof}

\begin{rem}
It is noted that by the cascaded structure of the local partial state observer \eqref{eq:z_i} and the consensus network \eqref{eq:x_i}, analysis for the case when there are disturbances in the system and/or noises in the output is not difficult.
In this case, it can be shown that the estimation error cannot be made arbitrarily small regardless how large the gains are. \QEDB
\end{rem}

\subsection{Special Case: Lyapunov Stable System}

In this subsection, under the additional assumption that the system \eqref{eq:sys} is Lyapunov stable, we show that the result of Theorem \ref{thm:suff} can be extended to a global result so that the initial condition can have any value and the gains $\kappa$ and $\gamma$ can be just positive numbers.
In particular, if there exists $P>0$ such that $PA+A^TP \le 0$, then the median solver \eqref{eq:x_i} can be modified as
\begin{align} \label{eq:x_ii}
\dot {\hat x}_i &= A \hat x_i + B u + \kappa {\mathcal V}\sqrt{\bar P^{-1}}{\mathcal W} \sum_{l = 1}^n s_i^l \text{sgn}(\bar w_l^T\VV_iz_i - \bar w_l^T\hat{x}_i) v_l \nonumber \\
&\quad\quad\quad\quad\quad\quad\quad\quad\quad\quad\quad\quad + \kappa\gamma \sum_{j \in \mathcal{N}_i}(\hat{x}_j - \hat{x}_i)
\end{align}
where $\bar P := {\mathcal V}^T P {\mathcal V}$ and $\bar w_l^T$ is the $l$-th row of the matrix $\sqrt{\bar P}{\mathcal W}$.
Now, the following holds.

\begin{thm}\label{thm:marg}
When the system \eqref{eq:sys} is Lyapunov stable, the modified median solver \eqref{eq:x_ii} works with \eqref{eq:z_i} under Assumptions \ref{assum:q}, \ref{assum:red}, \ref{assum:graph}, and \ref{assum:class}, with any $\kappa > 0$, $\gamma > 0$, and ${\rm col}(x(0), z_1(0), \hat{x}_1(0), \dots, z_N(0), \hat{x}_N(0)) \in \mathbb{R}^{n + \sum_{i=1}^N (o_i + n)}$.
In particular, the solution to \eqref{eq:sys}, \eqref{eq:z_i}, and \eqref{eq:x_ii} exists for all $t \ge 0$, and satisfies
\begin{align*}
\limsup_{t \to \infty} \|\hat{x}_i(t) - x(t)\| \le \frac{(Nn^2 + \sqrt{n})\sqrt{N}}{\gamma\lambda_2(\mathcal{L})}\left\|\mathcal{V} \sqrt{\bar P^{-1}}\right\|
\end{align*}
for all $i \in \mathcal{N}$. \QEDB
\end{thm}

\begin{IEEEproof}
Let $\bar x_i = \sqrt{\bar P}{\mathcal W}(\hat x_i - x)$ and $p_i = {\rm col}(p_i^1, \dots, p_i^n) = \sqrt{\bar P}{\mathcal W}({\mathcal V}_iz_i - x)$.
Then, one can show that \eqref{eq:key} still holds with $\bar A = \sqrt{\bar P} {\mathcal W} A {\mathcal V} \sqrt{\bar P^{-1}}$, which satisfies $\bar A + \bar A^T \le 0$.

Now, the rest of the proof proceeds similarly to the proof of Theorem \ref{thm:suff}.
In particular, the inequality \eqref{eq:dyn_W} now becomes
\begin{align}\label{eq:dyn_W2}
\dot{W} &\le \kappa\sqrt{Nn} - \kappa\gamma\lambda_2(\mathcal{L})W.
\end{align}
By this, we have
$$\limsup_{t \to \infty} \|\tilde{x}(t)\| = \limsup_{t \to \infty} W(t) \le \frac{\sqrt{Nn}}{\gamma\lambda_2(\mathcal{L})}.$$
This means that, for any $\epsilon > 0$, one can find $T > 0$ such that
\begin{align}\label{eq:dyn_tilde3}
\|\tilde{x}(t)\| \le \frac{\sqrt{Nn}}{\gamma\lambda_2(\mathcal{L})} + \epsilon, \quad \forall t \ge T.
\end{align}
Now, assume without loss of generality that $T$ is large enough so that
\begin{align}\label{eq:dyn_p3}
|p_i^l(t)| \le \epsilon, \quad \forall t \ge T,
\end{align}
for any $(i,l)$ such that $s_i^l = 1$ and $a_i \equiv 0$.
Then, we have
$$|p_i^l(t) - (\fr_i^T \otimes e_l^T)\tilde{x}| \le |p_i^l(t)| + \|\tilde{x}(t)\| \le \frac{\sqrt{Nn}}{\gamma\lambda_2(\mathcal{L})} + 2\epsilon,$$
for all $t \ge T$ and for all such pairs $(i, l)$.

Now, the Claim in the proof of Theorem~\ref{thm:suff} can be read as it holds that 
$$\dot{V} \le -\frac{\kappa}{Nn\sqrt{n}} < 0$$
if $t \ge T$ and
$$\frac{\sqrt{Nn}}{\gamma\lambda_2(\mathcal{L})} + 2\epsilon < \frac{V}{Nn\sqrt{n}} \le \frac{\|\bxa\|_\infty}{Nn}.$$
Thus we obtain
$$\limsup_{t\to \infty} V(t) \le Nn\sqrt{n}\left(\frac{\sqrt{Nn}}{\gamma\lambda_2(\mathcal{L})} + 2\epsilon\right).$$
However, the choice of $\epsilon$ is arbitrary, and thus, we get
\begin{align*}
\limsup_{t\to \infty} \|\bar{x}_{\text{avg}}(t)\| = \limsup_{t\to \infty}V(t)  \le \frac{Nn^2\sqrt{N}}{\gamma\lambda_2(\mathcal{L})}.
\end{align*}
Since $\bar{x}_i(t) = \bar{x}_{\text{avg}}(t) + (\fr_i^T \otimes I_n)\tilde{x}(t)$, this concludes the proof.
\end{IEEEproof}

Since the steady-state error depends only on the parameter $\gamma$, by increasing $\gamma$ and decreasing $\kappa$, we can achieve arbitrary small steady-state error, while preserving the coupling gain $\kappa\gamma$ as a constant.
However, as one can find from the proof of Theorem~\ref{thm:marg}, if $\kappa$ is small, the convergence rate is also small.
One way of achieving both fast convergence and small steady-state error is again taking $\kappa$ and $\gamma$ sufficiently large.

Meanwhile, by considering Lyapunov stable systems the method to allow plug-and-play operation noted in the Introduction and the method to identify effective attacks become much easy.
In particular, if each agent knows the upper bound of the number of agents as $\overline{N}$, and if there is a prespecified goal of steady-state error as $\overline{s}$, then each agent can simply take the parameters as
$$\gamma =  \frac{\|\VV\sqrt{\bar{P}^{-1}}\|\sqrt{\overline{N}}}{4\overline{s}/(\overline{N}^2 - \overline{N})} \left(\overline{N}n^2 + \sqrt{n}\right), \quad \kappa = \frac{1}{\gamma},$$
to guarantee (since $\lambda_2(\mathcal{L}) \ge 4/(N^2-N)$ \cite{mohar1991eigenvalues})
\begin{align*}
\limsup_{t \to \infty} \|\hat{x}_i(t) - x(t)\| \le \overline{s}, \quad \forall i \in \mathcal{N},
\end{align*}
with a coupling gain $\kappa\gamma = 1$, even when some sensor banks or local observers intermittently join or leave the network as long as the proposed assumptions are maintained.
Now, if we additionally assume that the initial condition is in some compact set $K$ as in Theorem~\ref{thm:suff}, then each agent can calculate $T$ that guarantees \eqref{eq:dyn_tilde3} with the prespecified $\epsilon$.
Then, by each agent constructing their partial observer fast enough so that \eqref{eq:dyn_p3} is satisfied for the same $T$ when $s_i^l = 1$ and $a_i \equiv 0$, they can also calculate $T' > T$ such that 
$$V(t) \le Nn\sqrt{n}\left(\frac{\sqrt{Nn}}{\gamma\lambda_2(\mathcal{L})} + 2\epsilon\right)$$
for all $t \ge T'$.
Now, from the time after $T'$, each agent can, by comparing their partial estimates $z_i(t)$ with their resilient estimates $\WW_i\hat{x}_i(t)$, find whether their measurement is corrupted by the effective attack or not.
Same ideas also apply to the general case, however, it requires much more efforts.

\section{Simulation Results}

\begin{figure}[!h]
\begin{center}
\includegraphics[width=0.8\columnwidth]{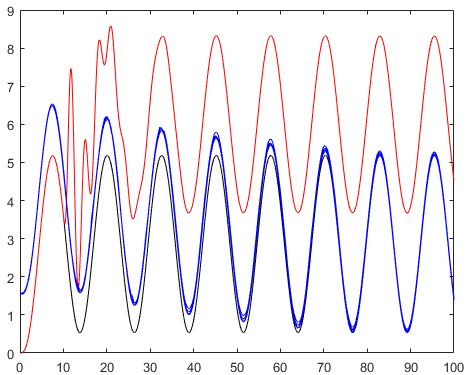}
\caption{Plot of the state trajectory $\theta_1(t) + \theta_2(t) + \theta_3(t)$ (black), its estimate obtained by agent $1$ (red), and its resilient estimate for all agents (blue).}
\label{fig:the1}
\end{center}
\vspace{-3mm}
\end{figure}

To verify the effectiveness of the proposed distributed resilient state estimator, simulation with a three inertia system is conducted.
Its dynamics \eqref{eq:sys} is determined by the matrices
\begin{align*}
A &= \begin{bmatrix} 0 & 1 & 0& 0& 0& 0\\ -\frac{k_1}{J_1} & -\frac{b_1}{J_1} & \frac{k_1}{J_1} & 0& 0& 0\\ 0& 0& 0& 1 & 0& 0\\ \frac{k_1}{J_2} & 0& -\frac{k_1 + k_2}{J_2} & -\frac{b_2}{J_2} & \frac{k_2}{J_2} &0 \\ 0& 0& 0& 0& 0& 1 \\ 0& 0& \frac{k_2}{J_3} & 0& -\frac{k_2}{J_3} & -\frac{b_3}{J_3}\end{bmatrix} \\
B &= \begin{bmatrix}  0\\ \frac{1}{J_1} \\ 0\\ 0\\ 0\\ 0\end{bmatrix}, \quad C = \begin{bmatrix} 1 &0 &0 &0 &0 &0 \\0 &0 & 1 &0 &0 &0 \\ 0& 0& 0& 0& 1 &0 \\ 1 & 0& -1 & 0& 0& 0\\ 0& 0& 1 & 0 & -1 & 0\end{bmatrix}
\end{align*}
where $J_1 = J_2 = J_3 = 0.01 \text{ kg}\cdot\text{m}^2$, $b_1 = b_2 = b_3 = 0.007 \text{ N}\cdot\text{m}/(\text{rad}/\text{s})$, and $k_1 = k_2 = 1.37 \text{ N}\cdot\text{m}/\text{rad}$.
Here, the state variables are $x:= \begin{bmatrix} \theta_1 \,\, \dot{\theta}_1 \,\, \theta_2 \,\, \dot{\theta}_2 \,\, \theta_3 \,\, \dot{\theta}_3\end{bmatrix}^T$, the output measurements are $y:= \begin{bmatrix} \theta_1 \,\, \theta_2 \,\, \theta_3 \,\, \theta_1 - \theta_2 \,\, \theta_2 - \theta_3\end{bmatrix}^T$ where each sensor bank consists of a single sensor, and the system is being controlled by $u = 0.01\sin(0.5 t)$.
Note that the pair $(C, A)$ is $2$-redundant observable and satisfies Assumption~\ref{assum:class}.
The injection gains $L_i$ of the partial observer \eqref{eq:z_i} are chosen appropriately such that the eigenvalues of $\mathcal{W}_iA\mathcal{V}_i - L_iC_i\mathcal{V}_i$ are near $-1$.
It is assumed that five agents are connected through the ring network, and $\kappa = 0.5$ and $\gamma = 2$ are used to construct \eqref{eq:x_i}.
Measurement data injection attack is applied to the first sensor as $a_1(t) = \pi/3$ for $t \ge 10$.
Fig.~\ref{fig:the1} shows state trajectory $\theta_1(t) + \theta_2(t) + \theta_3(t)$, its estimate obtained by agent~$1$, and its resilient estimate for all agents, which are obtained by the proposed network.\footnote{Note that the attack in this simulation is the so-called `zero-dynamics' attack, and thus, agent $1$ could not identify whether it is compromised or not, by only observing its own measurement.}
It demonstrates the attack-resilient property of our estimation algorithm.
Final emphasis is made that under the given network structure it is impossible for agent~$1$ to obtain resilient estimates corresponding to the subspace $\text{span}\{\begin{bmatrix} 1 \,\, 0 \,\, 1 \,\, 0 \,\, 1 \,\, 0 \end{bmatrix}^T, \begin{bmatrix} 0 \,\, 1 \,\, 0 \,\, 1 \,\ 0 \,\, 1\end{bmatrix}^T\}$ without a distributed majority voting, i.e., with only the collected measurements of its neighbors.

\section{Conclusion}\label{sec:con}
Under the assumption of $2q$-redundant observability and the connectivity of the network, a scheme of fully distributed resilient state estimation is proposed for a class of LTI systems having a uniform basis for unobservable subspaces of individual outputs.
Once the resilient estimate of the state is obtained, the attack is also identified by comparing the partial estimate $z_i$ of the local observer and the resilient estimate ${\mathcal W}_i \hat x_i$.
Since particular initialization is not needed for the algorithm \eqref{eq:z_i} and \eqref{eq:x_i} as seen in Theorem \ref{thm:suff}, the proposed scheme is suitable for the plug-and-play operation; that is, as long as the $2q$-redundant observability and the connectivity of the network is maintained, agents can freely join and leave the network during the operation.
The proposed scheme is fully distributed in that each local agent estimates the observable part of the state only, and the unobservable part is provided from the neighbors.
During this process, compromised data by the attacker are effectively rejected in a distributed way by the proposed distributed median solver.

\appendix

\subsection{Proof of Theorem~\ref{thm:sup_med}}\label{app:proof_dms}

The proof of Theorem~\ref{thm:sup_med} follows from the proof of Theorem~\ref{thm:marg}, by noting that, in this special case, we have $n = 1$, $A = 0$, $B = 0$, $m_i = 1$ for all $i \in \mathcal{N}$, $\kappa =1$, and $P = 1$.
In particular, we have $\dot{W} \le \sqrt{N} - \gamma \lambda_2(\mathcal{L})W$, and thus
\begin{align*}
\limsup_{t\to\infty} \|\tilde{x}(t)\| = \limsup_{t \to \infty} W(t) \le \sqrt{N}/(\gamma\lambda_2(\mathcal{L})).
\end{align*}
This means that, for any $\epsilon > 0$,  one can find $T > 0$ such that
$$\|\tilde{x}(t)\| \le \sqrt{N}/(\gamma\lambda_2(\mathcal{L})) + \epsilon, \quad \forall t \ge T.$$
Let $\underline{z}$ and $\overline{z}$ be such that $\mathcal{M}_{\mathcal{Z}}^{\mathcal{S}} = [\underline{z}, \overline{z}]$ and define
$$V(x_\avg) := \|x_\avg\|_{\mathcal{M}_\mathcal{Z}^{\mathcal{S}}} = \begin{cases} x_{\text{avg}} - \overline{z}, &\mbox{ if } x_{\text{avg}} > \overline{z}, \\ 0, &\mbox{ if } x_{\text{avg}} \in [\underline{z}, \overline{z}], \\ \underline{z} - x_{\text{avg}}, &\mbox{ if } x_{\text{avg}} < \underline{z}.\end{cases}$$
Now, suppose that $t \ge T$ and $x_{\avg}(t) > \overline{z} + \sqrt{N}/(\gamma\lambda_2(\mathcal{L})) + \epsilon$.
Then, $x_\avg(t) + \fr_i^T \tilde x(t) > \overline{z}$, which in turn implies that 
$$| \{ i \in \NN : x_\avg(t)+\fr_i^T \tilde x(t) > z_i \text{ and } s_i = 1\} | \ge \left\lceil (S+1)/2 \right\rceil \!.$$
Therefore, we get
\begin{align*}
\dot{V} &= \dot{x}_{\text{avg}}(t) = (1/N)\begin{matrix}\sum_{i=1}^N \end{matrix}s_i\text{sgn}(z_i - x_{\text{avg}}(t) - \fr_i^T\tilde{x}(t)) \\
&\le (1/N)\left(-\left\lceil(S+1)/2\right\rceil + \left\lfloor (S-1)/2\right\rfloor \right) \le -1/N < 0.
\end{align*}
Similarly, if $t \ge T$ and $x_{\text{avg}}(t) < \underline{z} - \sqrt{N}/(\gamma\lambda_2(\mathcal{L})) - \epsilon$, then we again get $\dot{V} \le - 1/ N < 0$.
Therefore, we obtain
$$\limsup_{t \to \infty} \|x_{\text{avg}}(t)\|_{\mathcal{M}_\mathcal{Z}^{\mathcal{S}}} \le \sqrt{N}/(\gamma\lambda_2(\mathcal{L})) + \epsilon.$$
However, the choice of $\epsilon$ is arbitrary, and thus, we get
$$\limsup_{t \to \infty} \|x_{\text{avg}}(t)\|_{\mathcal{M}_\mathcal{Z}^{\mathcal{S}}} \le \sqrt{N}/(\gamma\lambda_2(\mathcal{L})).$$
Since $x_i(t) = x_\avg(t) + \fr_i^T \tilde x(t)$, this concludes the proof.

\subsection{Illustration of the Applicable System Class}\label{app:ASC}

We inspect Assumption~\ref{assum:class} in the coordinates where $A$ has the real Jordan form (see \cite{RealJordan}) without loss of generality.
For simplicity, let us first consider the case when all eigenvalues of $A$ are real.
\begin{enumerate}

\item If $(C_i,A)$ is observable for all $i \in \NN$, Assumption \ref{assum:class} holds with any basis because all $\UU_i$'s are $\{0\}$.
This is the class of systems considered in \cite{jeon2016resilient}.

\item If $A$ has distinct eigenvalues, then the Jordan form is a diagonal matrix.
In this case, all the eigenvectors consist of elementary vectors $e_l$.
Therefore, Assumption \ref{assum:class} holds with $v_l = e_l$, $l = 1, \dots, n$.
The class of systems studied in \cite{mitra2016secure} belongs to this case.

\item More generally, if the characteristic polynomial of $A$ is the same as the minimal polynomial of $A$ (or, equivalently, each distinct eigenvalue of $A$ has only one Jordan block), Assumption \ref{assum:class} holds with $v_l = e_l$, $l = 1, \dots, n$.
To see this, without loss of generality suppose that $A$ is a single Jordan block.
Then, $\UU_i$ is $\{ 0\}$ when the first column of $C_i$ is non-zero, and, if the first $k$ columns of $C_i$ are all zero, then $\UU_i = \spn\{ e_{1}, \dots, e_{k} \}$ \cite[Sec.~6.5]{chen1998linear}.

\item Even in the case when there is more than one Jordan block for an eigenvalue $\lambda$, there are cases where Assumption \ref{assum:class} holds.
For example, consider
$$A = \left[\begin{smallmatrix} \lambda & 0 & 0 \\ 0 & \lambda & 1 \\ 0 & 0 & \lambda \end{smallmatrix}\right] \quad \text{and} \quad C = \left[\begin{smallmatrix} 1 & 1 & * \\ 1 & -1 & * \\ 2 & 2 & * \end{smallmatrix}\right]$$
and suppose that each sensor bank consists of only one sensor (so that $C_i$, $i=1,2,3$, is a single row vector).
In this case, Assumption \ref{assum:class} holds with $v_1 = [1,1, 0]^T$, $v_2 = [1,-1,0 ]^T$, and $v_3 = [0,0,1]^T$ because $\UU_1 = \UU_3 = \spn\{v_2\}$ and $\UU_2 = \spn\{v_1\}$.
However, if $C_3 = [2,2,*]$ is replaced by $[1,2,*]$, then $\UU_3 = \spn\{[2,-1, 0]^T\}$ so that there is no basis with which Assumption \ref{assum:class} holds.

\end{enumerate}

This observation indicates that indeed the pathological case may appear only when the characteristic polynomial of $A$ is not minimal, as noted in Section~\ref{sec:ACS}.

The above discussion can be extended to the case when $A$ has complex eigenvalues.
\begin{enumerate}
\setcounter{enumi}{4}

\item Same as the item 1) above.

\item Suppose that $A = \text{diag}(\Lambda_1,\dots,\Lambda_D)$ where $\Lambda_d$ is either $\lambda_d$ for real $\lambda_d$ or $\col([\alpha, -\beta],[\beta,\alpha])$ for $\lambda_d = \alpha+j\beta$ with $\beta \not = 0$, and $D$ is the number of distinct eigenvalues of $A$ when the complex conjugate eigenvalues are counted as one.
Then, it is seen that the unobservable subspace $\UU_i$ is a span of elementary vectors, by recalling that a complex mode $\lambda$ is unobservable if and only if its conjugate mode $\bar \lambda$ is unobservable for real matrix $A$.

\item The argument is the same as the item 3) above except that, for example, if 
$$A = \left[\begin{smallmatrix} \alpha & -\beta & 1 & 0 \\ \beta & \alpha & 0 & 1 \\ 0 & 0 & \alpha & -\beta \\ 0 & 0 & \beta & \alpha \end{smallmatrix}\right], \qquad \beta \not = 0,$$
the unobservable subspace $\UU_i$ is either $\{0\}$, $\spn\{e_1,e_2\}$, or $\spn\{e_1,\dots,e_4\}$ depending on whether the submatrix of the first two-columns of $C_i$ is non-zero, the first two-columns are zero but the submatrix of the last two-columns is non-zero, or all columns are zero, respectively.

\item Similar to the item 4) above, consider the case when
$$A = \left[\begin{smallmatrix} 0 & 1 & 0 & 0 \\ -1 & 0 & 0 & 0 \\ 0 & 0 & 0 & 1 \\ 0 & 0 & -1 & 0 \end{smallmatrix}\right] \; \text{with} \; C = \left[\begin{smallmatrix} -1 & 0 & 0 & 1 \\ 1 & 1 & 0 & 0 \\ 0 & 1 & 1 & 0 \\ 1 & -1 & 0 & 0 \\ -1 & 1 & 1 & 1 \\ 1 & 0 & 0 & 0 \end{smallmatrix}\right]$$
where each sensor bank consists of a single sensor.
Then, Assumption \ref{assum:class} holds with $v_1 = e_3$, $v_2 = e_4$, $v_3 = [0, 1, -1, 0]^T$, $v_4 = [1, 0, 0, 1]^T$.
Indeed, one can verify that $\UU_1 = \UU_3 = \UU_5 = \spn\{v_3,v_4\}$ and $\UU_2 = \UU_4 = \UU_6 = \spn\{v_1,v_2\}$.
However, if $C_1 = [-1,0,0,1]$ is replaced by $[2,0,0,1]$ for example, Assumption \ref{assum:class} does not hold anymore since $\UU_1 = \spn\{[1,0,0,-2]^T, [0,1,2,0]^T\}$.

\end{enumerate}

\end{document}